\pdfoutput=1

\documentclass[11pt,twoside,a4paper,cmspaper,final,collab]{cms-tdr}
\def\svnVersion{5e3c674-D}\def\svnDate{2020/01/23}\def\cmsCernNoTag{CERN-EP-2019-224}\def\cmsCernDate{\today}\def\cmsMessage{"Published in Physics Letters B as \href{http://dx.doi.org/10.1016/j.physletb.2020.135203}{\doi{10.1016/j.physletb.2020.135203}.}"}

\begin{document}\cmsNoteHeader{BPH-19-002}

\hyphenation{had-ron-i-za-tion}
\hyphenation{cal-or-i-me-ter}
\hyphenation{de-vices}
\renewcommand{\arraystretch}{1.1}
\newcommand{\ra}{\to}
\newcommand{\lam}{\Lambda}
\newcommand{\lb}{\Lambda_{\mathrm{b}}^0}
\newcommand{\jpsi}{\JPsi}
\newcommand{\pim}{\Pgpm}
\newcommand{\pip}{\Pgpp}
\newcommand{\mM}{\Pgmm}
\newcommand{\mP}{\Pgmp}
\newcommand{\psI}{\Pgy}
\newcommand{\kaon}      {\ensuremath{\PK}\xspace}
\newcommand{\Kp}      {\ensuremath{\PKp}\xspace}
\newcommand{\Km}      {\ensuremath{\PKm}\xspace}
\newcommand{\Kz}    {\ensuremath{\PKz}\xspace}
\newcommand{\bigcell}[2]{\begin{tabular}{@{}#1@{}}#2\end{tabular}}
\newlength\cmsTabSkip\setlength{\cmsTabSkip}{1ex}

\providecommand{\cmsTable}[1]{\resizebox{\columnwidth}{!}{#1}}
\newlength\cmsFigWidth
\ifthenelse{\boolean{cms@external}}{\setlength\cmsFigWidth{0.85\columnwidth}}{\setlength\cmsFigWidth{0.4\textwidth}}
\ifthenelse{\boolean{cms@external}}{\providecommand{\cmsLeft}{upper\xspace}}{\providecommand{\cmsLeft}{left\xspace}}
\ifthenelse{\boolean{cms@external}}{\providecommand{\cmsRight}{lower\xspace}}{\providecommand{\cmsRight}{right\xspace}}
\cmsNoteHeader{BPH-19-002}

\title{Observation of the $\lb \to \mathrm{J}/\psi \Lambda \PGf$ decay in proton-proton collisions at $\sqrt{s}=13\TeV$}

\date{\today}
\hyphenation{corres-ponding}
\abstract{
  The observation of the $\lb \ra \jpsi \lam \PGf$ decay is reported
  using proton-proton collision data collected at $\sqrt{s} = 13\TeV$ by the CMS experiment at the LHC in 2018,
  corresponding to an integrated luminosity of 60\fbinv.
  The ratio of the branching fractions
  $\mathcal{B}(\lb \ra \jpsi \lam \PGf)/\mathcal{B}(\lb\ra \psI \lam)$
  is measured to be $(8.26\pm0.90\stat \pm0.68\syst\pm0.11(\mathcal{B}))\times 10^{-2}$, where the first uncertainty is statistical, the second is systematic, and the last uncertainty reflects the
  uncertainties in the world-average branching fractions of $\PGf$ and $\psI$ decays to the reconstructed final states.
}
\hypersetup{
pdfauthor={CMS Collaboration},
pdftitle={Observation of the Lambda b 0 to J/psi Lambda phi decay in proton-proton collisions at sqrt(s) = 13 TeV},
pdfsubject={CMS},
pdfkeywords={CMS, b physics, heavy-flavor spectroscopy, computing}}

\maketitle
\section{Introduction}
Studies of \PQb baryon decays are of great importance for probing the dynamics of heavy-flavor decay processes. Since the observation of the lightest \PQb baryon $\lb$ by the UA1 Collaboration~\cite{ua1Lb} at the CERN S$\Pp\Pap$S, followed by extensive studies at the Fermilab Tevatron by the CDF~\cite{Abe:1992va,Abe:1996df,Abe:1996tr,Acosta:2005ab,Abulencia:2006df,Abulencia:2006dr,Aaltonen:2008eu,Aaltonen:2009zn,Aaltonen:2011qs,CDF:2011aa} and \DZERO~\cite{Abazov:2004bn,Abazov:2007sf,Abazov:2007al,Abazov:2011wt,Abazov:2012iy,Abazov:2015vea} Collaborations, the ATLAS, CMS, and LHCb experiments have accomplished numerous $\lb$ baryon studies, made possible by the large production cross section of $\bbbar$ pairs at the CERN LHC. Among these studies are precision mass measurements of the ground and excited states~\cite{LbMass,LbExc}, as well as lifetime and polarization measurements~\cite{AtlasLb, LbLTLHCb, LbPCMS, LbLTCMS}. Most of these studies have been performed in the $\lb\ra \jpsi\lam$ decay channel. Recently, an observation of the $\lb$ baryon decay to an excited charmonium state $\lb\ra \psI\lam$ has been reported by the ATLAS Collaboration~\cite{AtlasPsi}, while the LHCb Collaboration observed other, higher-multiplicity decays involving charmonium states~\cite{LHCbMult,Aaij:2019zkm}. Decays of the $\lb$ baryon also proved to be a rich source of exotic spectroscopy, as has been demonstrated by the observation by LHCb~\cite{lhcb_penta,Aaij:2019vzc} of new pentaquark states $\mathrm{P}_\PQc(4312)^+$, $\mathrm{P}_\PQc(4380)^+$, and $\mathrm{P}_\PQc(4450)^+$ in the invariant mass distribution of the $\jpsi\Pp$ system produced in the $\lb\ra \jpsi\Pp\Km$ decay. Further studies of the $\lb$ baryon decay modes involving charmonium states may shed light on the strong interaction processes in hadronic decays of \PQb baryons and on the production of exotic multiquark states. 

This Letter reports the observation of the $\lb\ra \jpsi\lam\PGf$ decay mode and the measurement of the branching fraction ratio $\mathcal{B}(\lb \ra \jpsi\lam \PGf)/\mathcal{B}(\lb \ra \psI \lam)$, by the CMS experiment. Here and thereafter, $\PGf$ refers to the $\PGf(1020)$ meson. The $\jpsi$, $\lam$, $\PGf$, and $\psI$ candidates are reconstructed in $\mP\mM$, $\Pp\pim$, $\Kp\Km$, and $\jpsi\pip\pim$ final states, respectively. The $\lb\ra \psI\lam \ra \jpsi\pip\pim\Pp\pim \ra \mP\mM\pip\pim\Pp\pim$ decay is used as the normalization channel, owing to its similar decay topology.

The branching fraction ratio $\mathcal{B}(\lb \ra \jpsi\lam \PGf)/\mathcal{B}(\lb \ra \psI \lam)$ is measured as:
\begin{linenomath}
\ifthenelse{\boolean{cms@external}}
{ 
\begin{multline}
  \dfrac{\mathcal{B}(\lb\ra \jpsi\lam\PGf)}{\mathcal{B}(\lb\ra \psI\lam)}  = \dfrac{N(\lb\ra \jpsi\lam\PGf)}{N(\lb\ra \psI\lam)}\\
  \times\dfrac{\epsilon(\lb\ra \psI\lam)\ \mathcal{B}(\psI\ra\jpsi\pip\pim)}{\epsilon(\lb\ra \jpsi\lam\PGf)\ \mathcal{B}(\PGf\ra\PKp\PKm)},\label{BFeq}
\end{multline}
}
{
\begin{equation}
  \dfrac{\mathcal{B}(\lb\ra \jpsi\lam\PGf)}{\mathcal{B}(\lb\ra \psI\lam)}  = \dfrac{N(\lb\ra \jpsi\lam\PGf)\ \epsilon(\lb\ra \psI\lam)\ \mathcal{B}(\psI\ra\jpsi\pip\pim)}{N(\lb\ra \psI\lam)\ \epsilon(\lb\ra \jpsi\lam\PGf)\ \mathcal{B}(\PGf\ra\PKp\PKm)},\label{BFeq}
\end{equation}
}
\end{linenomath}
where $N(\lb\ra \jpsi\lam\PGf)$ and $N(\lb\ra \psI\lam)$ are the measured $\lb$ yields for the signal and normalization channels, respectively. The terms $\epsilon(\lb\ra \jpsi\lam\PGf)$ and $\epsilon(\lb\ra \psI\lam)$ are the respective overall efficiencies that include the detector acceptance and the reconstruction efficiency. The branching fractions $\mathcal{B}(\psI\ra\jpsi\pip\pim)$ and $\mathcal{B}(\PGf\ra\PKp\PKm)$ are taken from the Particle Data Group (PDG)~\cite{PDG2018}.

The $\lb\ra \jpsi\lam\PGf$ decay is expected to proceed via the $\PQb\ra\ccbar\cPqs$ process, similarly to the $\lb\ra \jpsi\lam$ decay, but requires an additional $\cPqs\cPaqs$ pair. Consequently, the measurement of its branching fraction could enhance the understanding of the final-state strong interactions in \PQb baryon decays and test heavy-quark effective theory~\cite{hqet}. In addition, the $\lb\ra \jpsi\lam\PGf$ decay is a baryonic analog of the $\PB^{+} \to \jpsi \PGf \PK^{+}$ decay, where a rich resonant structure in the $\jpsi \PGf$ system has been observed by several experiments~\cite{CDF4140,CMS4140,D04140,LHCb4140}. Therefore, detailed studies of the $\jpsi \PGf$ spectrum produced in baryonic decays may provide an important test for the production of these states. Recently, the existence of a hidden-charm pentaquark spectra was predicted for the $\jpsi\lam$ final state~\cite{PhysRevD.100.016014}, which can be investigated in the $\lb\ra \jpsi\lam\PGf$ decay, once a sufficient number of signal events is accumulated.

\hyphenation{dia-meter}
\section{The CMS detector}
The central feature of the CMS apparatus~\cite{d3} is a superconducting solenoid of 6\unit{m} internal diameter, providing a magnetic field of 3.8\unit{T}. Within the solenoid volume are a silicon pixel and strip tracker, a lead tungstate crystal electromagnetic calorimeter, and a brass and scintillator hadron calorimeter, each composed of a barrel and two endcap sections. Forward calorimeters extend the pseudorapidity ($\eta$) coverage provided by the barrel and endcap detectors. Muons are detected in gas-ionization chambers embedded in the steel flux-return yoke outside the solenoid. The main subdetectors used for the present analysis are the silicon tracker and the muon system.

The silicon tracker measures charged particles within the range $\abs{\eta} < 2.5$. During the LHC running period when the data used in this Letter were recorded, the silicon tracker consisted of 1856 silicon pixel and 15\,148 silicon strip detector modules. For nonisolated particles with transverse momentum $1 < \pt < 10\GeV$ and $\abs{\eta} < 1.4$, the track resolution is typically 1.5\% in \pt.

Muons are measured within $\abs{\eta} < 2.4$, with detection planes made using three technologies: drift tubes, cathode strip chambers, and resistive-plate chambers. Tracks in the muon system are matched to those measured in the silicon tracker. The relative $\pt$ resolution is measured to be in the range 0.8--3.0$\%$ for muons with $\pt < 10\GeV$ used in this analysis, depending on the muon $\abs{\eta}$~\cite{CMS-MUO-16-001}.

Events of interest are selected using a two-tiered trigger system~\cite{Khachatryan:2016bia}. The first level (L1), composed of custom hardware processors, uses information from the calorimeters and muon detectors to select events at a rate up to 100$\unit{kHz}$ within a fixed time interval of less than 4$\mus$. The second level, known as the high-level trigger (HLT), consists of a farm of processors running a version of the full event reconstruction software optimized for fast processing, and reduces the event rate to around 1$\unit{kHz}$ before data storage.

A more detailed description of the CMS detector, together with a definition of the coordinate system used and the relevant kinematic variables, can be found in Ref.~\cite{d3}.

\section{Data sample and event selection} \label{evtsel}
The analysis described in this Letter is based on a data sample of proton-proton collisions at a center-of-mass energy of 13\TeV, collected with the CMS detector in 2018 and corresponding to an integrated luminosity of 60\fbinv.

Data were recorded with a dedicated trigger, optimized for the selection of $\PQb$ hadrons decaying to $\JPsi(\PGmp\PGmm)$ and two additional tracks from the charged hadrons emerging from the decay. 
The L1 trigger requires two oppositely charged muons with \pt of at least 4\GeV, or two muons in the barrel region ($\abs{\eta} < 1.479$) without any \pt threshold.
At the HLT, a $\JPsi$ candidate decaying into a $\PGmp\PGmm$ pair displaced from the interaction point is required, along with at least two tracks consistent with the displaced vertex. Each muon $\pt$ is required to be at least 4\GeV, while the dimuon $\pt$ is required to exceed 6.9\GeV.  The $\JPsi$ candidates reconstructed from dimuons are required to have an invariant mass between 2.9 and 3.3\GeV. The three-dimensional distance of
closest approach of the two muons to each other is required to be less than 0.5\unit{cm}. The fitted dimuon vertex is required to have a transverse decay length significance
$L_{xy}(\jpsi)/\sigma_{L_{xy}(\jpsi)} > 3$, where $L_{xy}(\jpsi)$ and $\sigma_{L_{xy}(\jpsi)}$
are, respectively, the distance from the common vertex to the beam axis in the transverse plane and its uncertainty.
Finally, the dimuon vertex fit probability, calculated using the
$\chi^2$ and the number of degrees of freedom of the fit, is required to exceed 10$\%$, while
the angle $\alpha$ between the dimuon $\pt$ vector and the
direction connecting the beam axis and the dimuon vertex in the transverse plane is required to satisfy $\cos\alpha>0.9$.
Given the lack of a dedicated kaon identification, the two additional tracks are assigned a kaon mass hypothesis and required to have $\pt> 0.8\GeV$, $\abs{\eta} < 2.5$, and an invariant mass in a range of 0.95--1.30\GeV.

In the subsequent offline analysis we follow closely the selection of Ref.~\cite{BPH-18-005}. The \pt threshold on the two muon candidates of 4\GeV and the requirement of $\abs{\eta} < 2.4$ are kept. Two oppositely charged muon candidates are paired and required to originate from a common vertex. The vertex requirements applied at the HLT are confirmed in the offline selection. Also both muon candidates must match those that triggered the event readout. Dimuon candidates with an invariant mass within 100$\MeV$, which corresponds to approximately four effective widths, around the $\JPsi$ meson mass $M_\JPsi^\text{PDG}$ are selected (hereafter, $M_\mathrm{X}^\text{PDG}$ denotes the world-average mass of hadron X~\cite{PDG2018}), and the $\pt$ of the $\JPsi$ meson is required to exceed $7\GeV$.

To reconstruct a $\lb$ candidate, the $\JPsi$ candidate is combined with two oppositely charged, high-purity~\cite{d1} tracks, assumed to be kaon candidates, and a $\lam$ candidate. The $\pt$ of the tracks is required to exceed $0.8\GeV$, and their invariant mass must satisfy $0.99<M(\Kp \Km)<1.05\GeV$. The $\lam$ candidates are formed from displaced two-prong vertices under the assumption of the $\lam  \to \Pp \Pgpm$ decay, as described in Ref.~\cite{trackreco}. Daughter particles of the $\lam$ candidate are refitted to a common vertex with their invariant mass constrained to $M_{\lam}^\text{PDG}$, and the vertex fit probability is required to exceed 1\%. The proton mass is assigned to the higher-momentum daughter track. To select the candidates in the $\lam$ signal region, the following additional requirement is applied: $\abs{M(\Pp\Pgpm)-M_{\lam}^\text{PDG}}<7.5\MeV$. The width of this window is chosen to correspond to approximately three times the effective width of the reconstructed $\lam$ candidates. In addition, the $\lam$ candidate is required to have a transverse momentum in excess of $1\GeV$. 

As the last step of the reconstruction, a fit to the common vertex of the $\lam$ candidate, the two kaon tracks, and the dimuon pair is performed, with the dimuon mass constrained to $M_\JPsi^\text{PDG}$; this vertex is referred to as the $\lb$ vertex. The kinematic vertex fit probability of the $\lb$ candidate is required to exceed 1\%. The selected candidates are required to have $\pt(\lb)>10\GeV$.

Multiple proton-proton interactions in the same or nearby beam crossing (pileup) are present in the data, with an average multiplicity of 32, resulting in multiple reconstructed vertices in an event. The vertex with the lowest three-dimensional angle between the line connecting this vertex with the $\lb$ vertex and the $\lb$ candidate momentum is chosen as the primary vertex (PV).  The following requirement is used to select $\lb$ candidates consistent with originating from the PV: $\cos\alpha (\lb,\text{PV})>0.99$, where $\alpha(\lb,\text{PV})$ is the two-dimensional angle in the transverse plane between the $\lb$ candidate momentum and the vector pointing from the PV to the $\lb$ vertex. The following requirement on the $\lb$ vertex displacement is also applied: $L_{xy}(\lb)/\sigma_{L_{xy}(\lb)} >3$, where $L_{xy}(\lb)$ is the distance between the primary and $\lb$ vertices in the transverse plane, and $\sigma_{L_{xy}(\lb)}$ is its uncertainty.

Candidate decays for the normalization channel $\lb \ra \psI \lam$, with $\psI \ra \jpsi \pip \pim$, are selected using the same reconstruction chain. Identical requirements are used to select the $\JPsi$ candidate, $\pip$ and $\pim$ tracks, and $\lam$ candidate. An additional requirement is placed on the $\jpsi \pip \pim$ invariant mass, $\abs{M(\jpsi \pip \pim)-M_{\psI}^\text{PDG}} < 15\MeV$, to select $\psI$ candidates, where this window corresponds to approximately three effective widths of a reconstructed $\psI$ candidate.

In case of multiple $\lb$ candidates per event, the one with the highest vertex fit probability is chosen for both the signal and normalization channels. There are 18.9 and 7.4$\%$ of events with two or more reconstructed candidates for signal and normalization channels, respectively. When there are two or more candidates in an event, the MC simulation predicts that the correct candidate is chosen $84 \pm 5$ and $93 \pm 13 \%$ of the time for the signal and normalization channels, respectively. 

To calculate the reconstruction efficiency, a study based on simulated signal events for both channels is performed. The events are generated with \PYTHIA  8.230~\cite{pythia}. The $\lb$ baryon decays are modeled with $\EVTGEN$~\cite{evtgen} v1.6.0 for both the $\lb \to \jpsi\lam \PGf$ and $\lb \to \psI \lam$ decay channels, following the three-body phase space model. The events are then passed through a detailed CMS detector simulation based on \GEANTfour~\cite{geant4}.

\section{Signal yield extraction}
\begin{figure*}[!htb]
\centering
\includegraphics[width=0.45\textwidth]{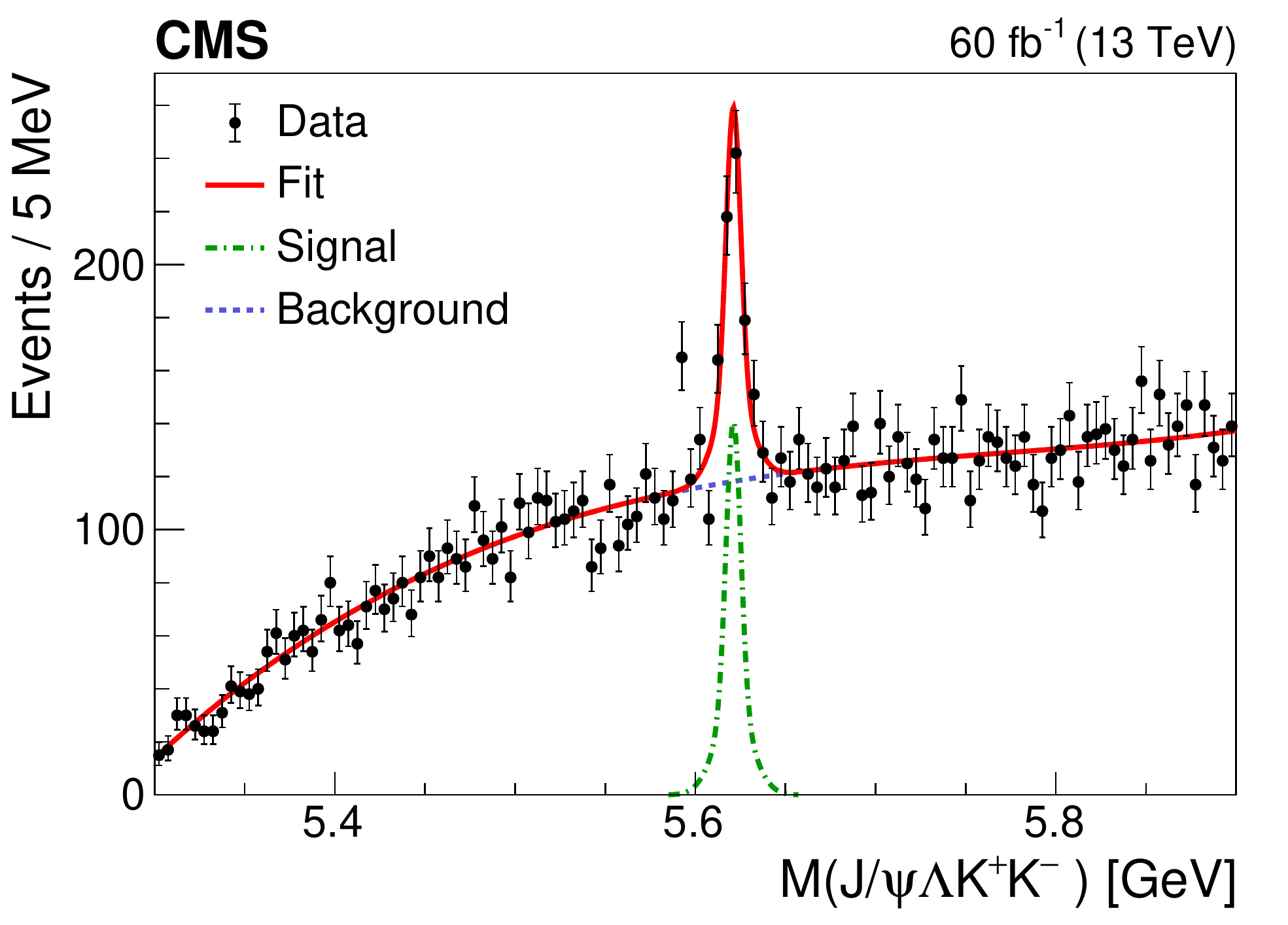}
\includegraphics[width=0.45\textwidth]{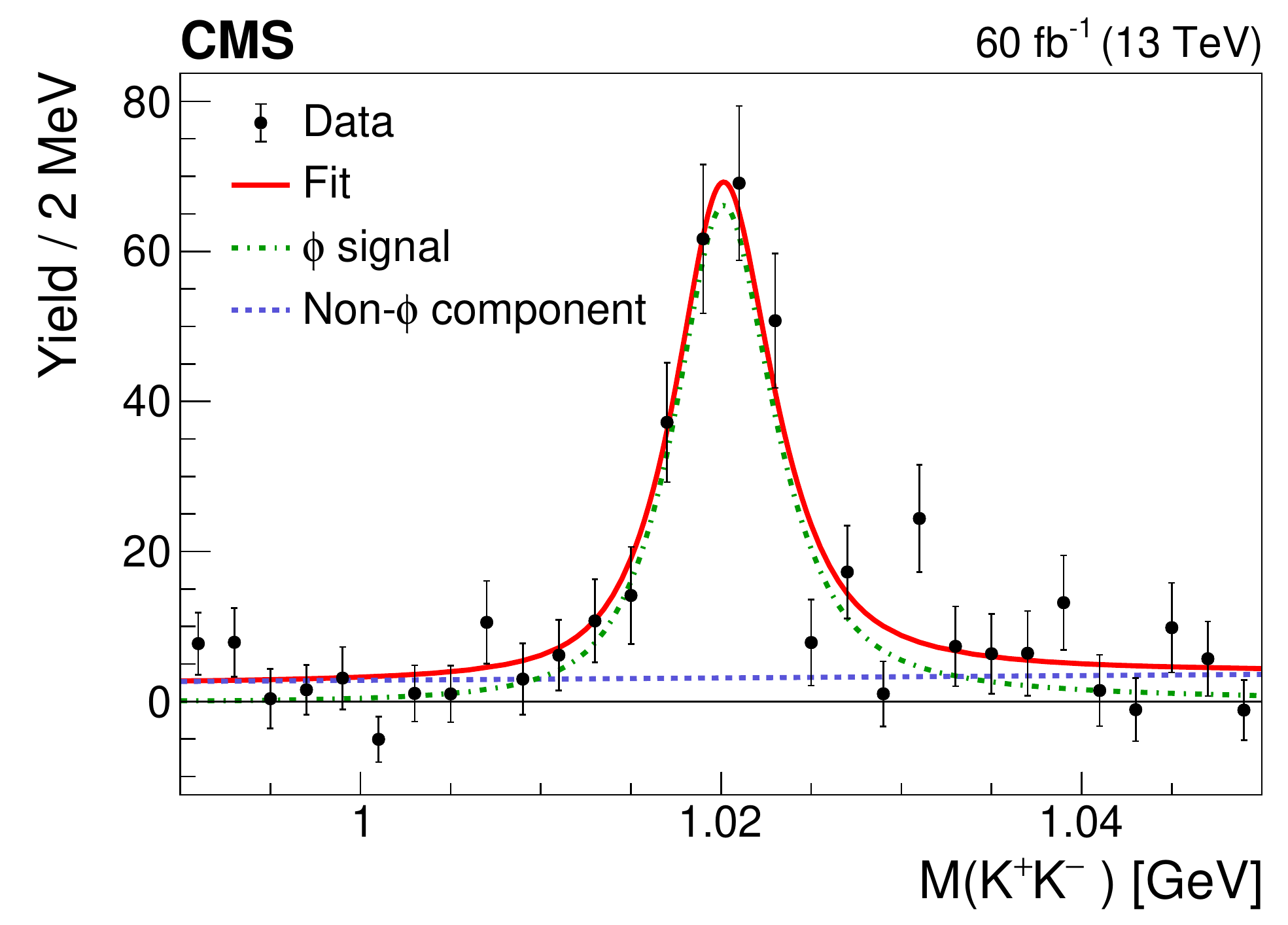}
\caption{
The invariant mass distributions of (left) $\jpsi\lam \Kp \Km$ and (right) background-subtracted $\Kp \Km$. The points are the data, with the vertical bars giving the statistical uncertainties, and the lines show the results of the fits described in the text.
}
\label{m0}
\end{figure*}

The invariant mass distribution of the $\lb \ra \jpsi \lam \Kp \Km$ candidates selected using the strategy described in the previous section is shown in Fig.~\ref{m0} (left). An unbinned, extended maximum-likelihood fit to a signal plus background hypothesis is performed on this observable and further mass distributions.

The signal is described by a double-Gaussian function with a floating common mean and total normalization,
while the two widths and the relative fraction of the two Gaussian functions are fixed to the values obtained from simulation.
The double-Gaussian function was chosen as a model that provides the best description of the simulated sample.
The background is parameterized by a third-order Bernstein polynomial.
The fit results in a signal yield of $380\pm 32$ events. The signal significance is calculated to be 9.7 standard deviations in the asymptotic approximation~\cite{CCGV}, using the profile likelihood ratio of the signal plus background over the background-only hypothesis as the test statistic. Including modeling uncertainties in the signal and background shapes (described in Section~\ref{systematic}) results in a reduction of the significance value to 9.4 standard deviations.

There is a bin with the yield significantly higher than the average background level in the left panel of Fig.~\ref{m0}, just below the signal $\lb$ peak. The local significance of the excess is estimated to be less than three standard deviations. Several cross-checks have been performed to investigate this enhancement. The $M(\lb \ra \jpsi \lam \Kp \Km)$ distribution with the requirement on the $\PGf$ candidates to have a mass within 10\MeV of the nominal value shows no significant excess in this bin. The statistical significance of the $\lb$ signal is 10.3 standard deviations in the asymptotic approximation. A statistically independent data set, collected in 2017, has been examined with the same selection, and no significant excess below the $\lb$ peak was observed. As a result of these cross-checks, we attribute the excess to a statistical fluctuation.

An unbinned likelihood fit to the $M(\lb \ra \jpsi \lam \Kp \Km)$ observable is employed to separate the signal and background components statistically, which is then used with the $\textit{sPlot}$ technique~\cite{sPlot} to obtain the $M(\Kp\Km)$ data distribution corresponding to signal $\lb \ra \jpsi \lam \Kp \Km$ decays. To extract the $\lb \ra \jpsi \lam \PGf$ decay yield, the background-subtracted $M(\Kp\Km)$ distribution is fitted with the convolution of a double-Gaussian and relativistic Breit--Wigner functions for the $\PGf$ signal and a first-order Bernstein polynomial for the nonresonant component. The natural width of the $\PGf$ meson is fixed to the world-average value~\cite{PDG2018}. It was checked that the natural width of the $\PGf$ meson obtained from the fit when it was allowed to float was consistent with the world-average value within the uncertainties. Both widths and the relative fraction of the two Gaussians are fixed to the values obtained from fitting the simulated signal sample. The fit results in a signal yield of $286 \pm 29$ events. The $M(\Kp\Km)$ invariant mass distribution, along with the result of the fit, are shown in Fig.~\ref{m0} (right). 

\begin{figure}[!htb]
\centering
\includegraphics[width=0.45\textwidth]{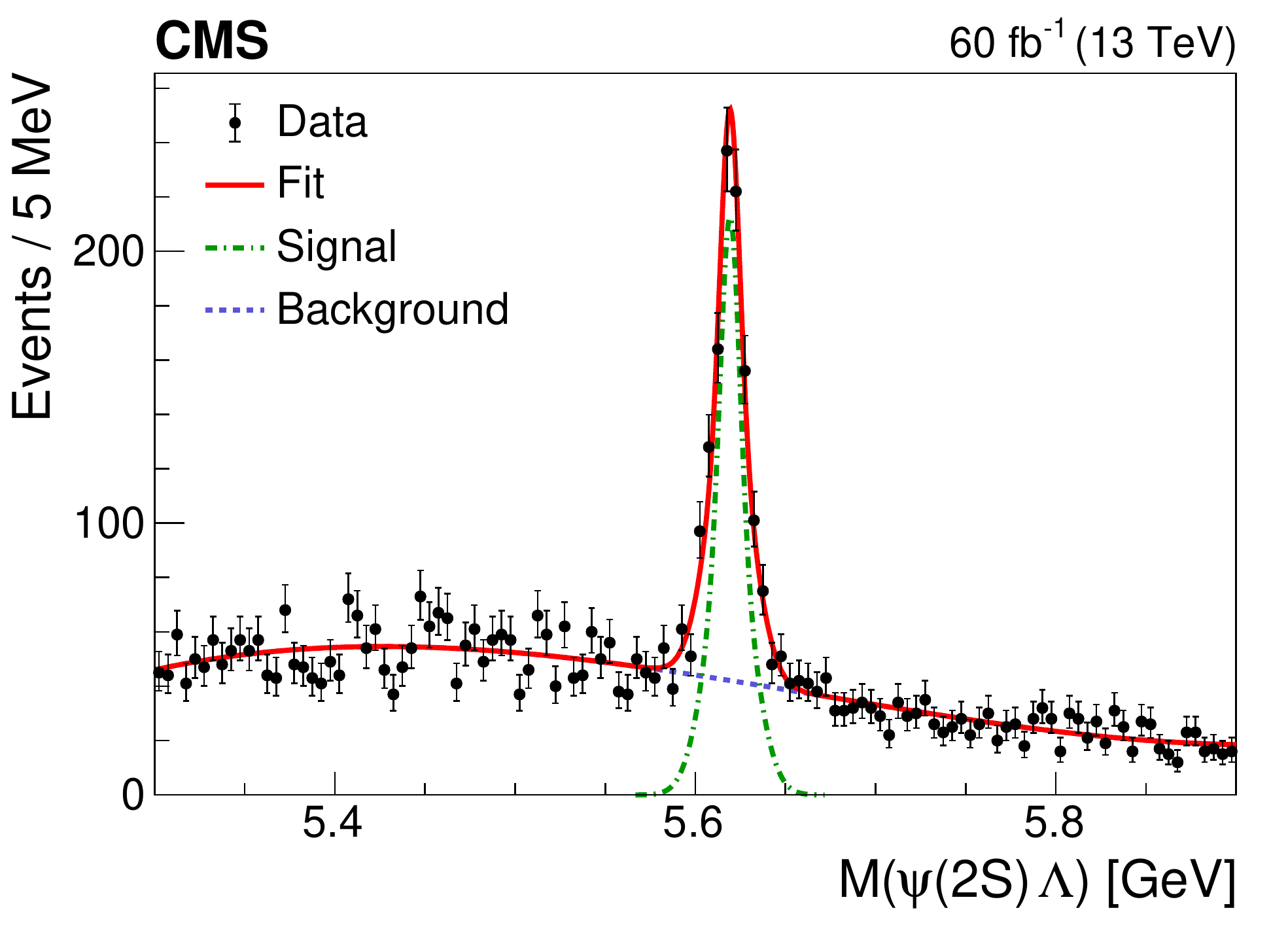}
\caption{
The invariant mass distribution of $\lb \ra \psI \lam$ candidates.
The points are the data and the lines give the result of the fit described in the text.}
\label{m2}
\end{figure}

Figure~\ref{m2} displays the invariant mass distribution of $\lb \ra \psI \lam$ candidates.
The points represent the data and the curve is the result of the fit.
The signal is described by a double-Gaussian function with floating common mean and 
total normalization, while the individual widths and the relative fraction of the two Gaussians are fixed from the fit to a simulated signal sample. The background is described by a third-order Bernstein polynomial function. The fit results in a signal yield of $884 \pm 37$ events. The non-$\psI$ contribution in the $\lb \ra \jpsi \pip \pim \lam$ signal was estimated to be negligible in the selected mass window $\abs{M(\jpsi \pip \pim)-M_{\psI}^\text{PDG}} < 15\MeV$.  

\section{Efficiency calculation}\label{EffSec}

The $\lb$ selection efficiencies in the signal and normalization channels are calculated as the ratio of the numbers of selected to generated events in simulated signal samples. The overall efficiency includes the trigger and reconstruction efficiencies and the detector acceptance. The efficiency in each channel is obtained using the simulated samples described in Section~\ref{evtsel}. The efficiency ratio, which is used in the branching fraction ratio measurement, is found to be ${\epsilon}(\lb \ra \psI \lam)/{\epsilon}(\lb \ra \jpsi\lam \PGf) =  0.363 \pm 0.011$, where the uncertainty is statistical only and accounts for the limited event counts in the corresponding simulated samples. The $\pt$ spectrum of pions from the $\psI \ra \jpsi \pip \pim$ decay in the normalizations channel is softer than the $\pt$ spectrum of kaons from the $\PGf \ra \Kp \Km$ decay in the signal channel, resulting in an efficiency ratio significantly below unity.

\section{Systematic uncertainties}\label{systematic}
{\tolerance=800
In this section we discuss various sources of systematic uncertainty contributing to the measurement of the ratio $\mathcal{B}(\lb \ra \jpsi\lam \PGf) / \mathcal{B}(\lb \ra \psI \lam)$, as defined in Eq.~(\ref{BFeq}).
\par}
Since both the $\lb \ra \jpsi\lam \PGf \ra \mP\mM\Pp\pim\Kp\Km$ and $\lb \ra \psI \lam \ra \mP\mM\Pp\pim\pip\pim$ decay modes have the same topology, the systematic uncertainties related to the muon and track reconstruction, as well as the trigger efficiency, mostly cancel in the ratio. 
To test this assumption, simulated samples were compared with background-subtracted data in a number of kinematic distributions. As a result of these studies, an additional systematic uncertainty is assigned to account for the observed difference between data and simulation in the $\lb$ rapidity distribution for the normalization channel, as well as for the difference in the two-body invariant mass distributions $M(\jpsi\lam)$, $M(\jpsi\PGf)$, and $M(\lam\PGf)$ in data and simulation for the signal channel. The latter discrepancy could be caused by a deviation from the pure phase space decay model used in the simulation due to contributions from intermediate resonant states; however, the statistical power of the present data set is insufficient to perform a more detailed investigation. To estimate this systematic uncertainty, the simulated samples were reweighted to match the distributions observed in data. The difference in the efficiency ratio before and after the reweighting is taken as the corresponding systematic uncertainty.

The systematic uncertainty related to the choice of the background model is estimated separately for the signal channel, normalization channel, and $\PGf \ra \Kp \Km$ decays. The variation of the background model includes Bernstein polynomials of second and fourth orders, independently for the signal and normalization channels, and an exponential function for the background in the $\PGf \ra \Kp \Km$ invariant mass distribution. For the signal channel, an additional background function with a threshold behavior is also tested: $(x-x_{0})^{\beta}$ multiplied by the Bernstein polynomials of first and second orders, where $x_0 = M_\JPsi^\text{PDG} + M_{\lam}^\text{PDG} + M_{\PGf}^\text{PDG}$ and the exponent $\beta$ is allowed to vary freely in the fit. In each case, the maximum deviation in the measured signal yield within the variations of the background model is used as the systematic uncertainty.

Another source of systematic uncertainty is the signal shape modeling in the $M(\jpsi\lam\Kp\Km)$, $M(\psI\lam)$, and $M(\Kp\Km)$ distributions. This uncertainty is estimated by using alternative signal models whose parameters were obtained by fitting the simulated invariant mass distributions. The variation of signal models includes a triple-Gaussian function and a sum of two Crystal Ball~\cite{CB1} functions for the $\lb \ra \jpsi\lam \Kp \Km$ invariant mass distribution; a sum of two Crystal Ball functions for the $\lb \ra \psI \lam$ channel; and a convolution of a double Crystal Ball~\cite{CB2} and relativistic Breit--Wigner functions for the $M(\Kp\Km)$ distribution. For each of the variations, the largest deviation in the measured signal yield is taken as the systematic uncertainty.

The next source of systematic uncertainty is the difference in the mass resolution of the $\lb$ and $\PGf$ peaks between data and simulation. To estimate this uncertainty, several variations were applied to the resolution functions in the $M(\jpsi\lam\Kp\Km)$ and $M(\psI\lam)$ distributions: only the ratio of the two Gaussian widths was fixed to the one measured in simulation instead of fixing both widths, as in the nominal fit.  For the $M(\Kp\Km)$ distribution, a fit with the fixed ratios of the two Gaussian widths and yields, as measured in simulation, is performed. In each case, the maximum variation in the measured $\lb$ yield is used as the systematic uncertainty. The difference between data and simulation in the measured $\lb$ mass resolution for the $\lb \ra \jpsi\lam \Kp \Km$ channel results in the largest systematic uncertainty. 

The statistical uncertainty in the efficiency ratio obtained from simulation is also considered as a source of systematic uncertainty.
Table~\ref{table_systematics} summarizes the individual sources of the systematic uncertainty, as well as the overall uncertainty obtained as a quadratic sum of the individual components.  

\begin{table*}[htb]
\centering
\topcaption {Summary of the relative systematic uncertainties in $\mathcal{B}(\lb \ra \jpsi\lam \PGf)/\mathcal{B}(\lb \ra \psI \lam)$.}
\begin{linenomath}
\ifthenelse{\boolean{cms@external}}
{
\label{table_systematics}
\begin{tabular}{lc}
\hline
Source  & Relative \\
&  uncertainty (\%)\\
\hline
Data/simulation difference in the $\lb$ rapidity and two-body mass distributions &0.1\\
Background model in the $M(\jpsi\lam \Kp \Km)$ distribution&0.6\\
Background model in the $M(\psI \lam)$ distribution&0.8 \\
Background model in the $M(\Kp \Km$) distribution&0.8\\
Signal model in the $M(\jpsi\lam \Kp \Km)$ distribution&0.8 \\
Signal model in the $M(\psI \lam)$ distribution & 1.1\\
Signal model in the $M(\Kp\Km)$ distribution  &0.5\\
Data/simulation difference in the $\lb$ resolution for the $\lb \ra \jpsi\lam \PGf$ decay& 6.6 \\
Data/simulation difference in the $\lb$ resolution for the $\lb \ra \psI \lam$ decay& 3.1 \\
Data/simulation difference in the $\PGf$ resolution& 1.4 \\
Finite size of simulated samples &2.9\\[\cmsTabSkip]
Total systematic uncertainty &8.2\\
\hline
\end{tabular} 
}
{
\cmsTable{
\label{table_systematics}
\begin{tabular}{lc}
\hline
Source  & Relative \\
&  uncertainty (\%)\\
\hline
Data/simulation difference in the $\lb$ rapidity and two-body mass distributions &0.1\\
Background model in the $M(\jpsi\lam \Kp \Km)$ distribution&0.6\\
Background model in the $M(\psI \lam)$ distribution&0.8 \\
Background model in the $M(\Kp \Km$) distribution&0.8\\
Signal model in the $M(\jpsi\lam \Kp \Km)$ distribution&0.8 \\
Signal model in the $M(\psI \lam)$ distribution & 1.1\\
Signal model in the $M(\Kp\Km)$ distribution  &0.5\\
Data/simulation difference in the $\lb$ resolution for the $\lb \ra \jpsi\lam \PGf$ decay& 6.6 \\
Data/simulation difference in the $\lb$ resolution for the $\lb \ra \psI \lam$ decay& 3.1 \\
Data/simulation difference in the $\PGf$ resolution& 1.4 \\
Finite size of simulated samples &2.9\\[\cmsTabSkip]
Total systematic uncertainty &8.2\\
\hline
\end{tabular} } 
}
\end{linenomath}
\end{table*}

\section{Measurement of the branching fraction ratio}
Using Eq.~(\ref{BFeq}), the signal and normalization channel yields $N(\lb \ra \jpsi\lam \PGf)= 286\pm29$ and $N(\lb \ra \psI \lam)=884\pm37$, the efficiency ratio described in Section~\ref{EffSec}, and
the PDG values of $\mathcal{B}(\psI\ra\jpsi\pip\pim) = 0.347 \pm 0.003$ and $\mathcal{B}(\PGf\ra\Kp\Km) = 0.492\pm0.005$, we measure the ratio $\mathcal{B}(\lb \ra \jpsi\lam \PGf)/\mathcal{B}(\lb \ra \psI \lam)$ to be $(8.26\pm0.90\stat \pm0.68\syst\pm0.11(\mathcal{B}))\times 10^{-2}$.
The first uncertainty is statistical, while the second is systematic (as described in Section~\ref{systematic}), and the third is due to the uncertainties in the branching fractions of the decays involved.

\section{Summary}
The observation of the $\lb \ra \jpsi\lam \PGf$ decay and the measurement of the branching fraction ratio $\mathcal{B}(\lb \ra \jpsi\lam \PGf)/\mathcal{B}(\lb \ra \psI \lam)$ is presented using a data sample of proton-proton collisions at $\sqrt{s} = 13\TeV$ collected in 2018 by the CMS experiment and corresponding to an integrated luminosity of 60\fbinv. The ratio $\mathcal{B}(\lb \ra \jpsi\lam \PGf)/\mathcal{B}(\lb \ra \psI \lam)$ is measured to be $(8.26\pm0.90\stat \pm0.68\syst\pm0.11(\mathcal{B}))\times 10^{-2}$, where the first uncertainty is statistical, the second is systematic, and the last uncertainty reflects the uncertainties in the world-average branching fractions of $\PGf$ and $\psI$ decays to the reconstructed final states. The observation of the $\lb \ra \jpsi\lam \PGf$ decay opens a window on future searches for new resonances in the $\jpsi\lam$ and $\jpsi\PGf$ mass spectra, once a sufficient number of signal events is observed.

\begin{acknowledgments}
We congratulate our colleagues in the CERN accelerator departments for the excellent performance of the LHC and thank the technical and administrative staffs at CERN and at other CMS institutes for their contributions to the success of the CMS effort. In addition, we gratefully acknowledge the computing centers and personnel of the Worldwide LHC Computing Grid for delivering so effectively the computing infrastructure essential to our analyses. Finally, we acknowledge the enduring support for the construction and operation of the LHC and the CMS detector provided by the following funding agencies: BMBWF and FWF (Austria); FNRS and FWO (Belgium); CNPq, CAPES, FAPERJ, FAPERGS, and FAPESP (Brazil); MES (Bulgaria); CERN; CAS, MoST, and NSFC (China); COLCIENCIAS (Colombia); MSES and CSF (Croatia); RPF (Cyprus); SENESCYT (Ecuador); MoER, ERC IUT, PUT and ERDF (Estonia); Academy of Finland, MEC, and HIP (Finland); CEA and CNRS/IN2P3 (France); BMBF, DFG, and HGF (Germany); GSRT (Greece); NKFIA (Hungary); DAE and DST (India); IPM (Iran); SFI (Ireland); INFN (Italy); MSIP and NRF (Republic of Korea); MES (Latvia); LAS (Lithuania); MOE and UM (Malaysia); BUAP, CINVESTAV, CONACYT, LNS, SEP, and UASLP-FAI (Mexico); MOS (Montenegro); MBIE (New Zealand); PAEC (Pakistan); MSHE and NSC (Poland); FCT (Portugal); JINR (Dubna); MON, RosAtom, RAS, RFBR, and NRC KI (Russia); MESTD (Serbia); SEIDI, CPAN, PCTI, and FEDER (Spain); MOSTR (Sri Lanka); Swiss Funding Agencies (Switzerland); MST (Taipei); ThEPCenter, IPST, STAR, and NSTDA (Thailand); TUBITAK and TAEK (Turkey); NASU (Ukraine); STFC (United Kingdom); DOE and NSF (USA).

\hyphenation{Rachada-pisek} Individuals have received support from the Marie-Curie program and the European Research Council and Horizon 2020 Grant, contract Nos.\ 675440, 752730, and 765710 (European Union); the Leventis Foundation; the A.P.\ Sloan Foundation; the Alexander von Humboldt Foundation; the Belgian Federal Science Policy Office; the Fonds pour la Formation \`a la Recherche dans l'Industrie et dans l'Agriculture (FRIA-Belgium); the Agentschap voor Innovatie door Wetenschap en Technologie (IWT-Belgium); the F.R.S.-FNRS and FWO (Belgium) under the ``Excellence of Science -- EOS" -- be.h project n.\ 30820817; the Beijing Municipal Science \& Technology Commission, No. Z181100004218003; the Ministry of Education, Youth and Sports (MEYS) of the Czech Republic; the Lend\"ulet (``Momentum") Program and the J\'anos Bolyai Research Scholarship of the Hungarian Academy of Sciences, the New National Excellence Program \'UNKP, the NKFIA research grants 123842, 123959, 124845, 124850, 125105, 128713, 128786, and 129058 (Hungary); the Council of Science and Industrial Research, India; the HOMING PLUS program of the Foundation for Polish Science, cofinanced from European Union, Regional Development Fund, the Mobility Plus program of the Ministry of Science and Higher Education, the National Science Center (Poland), contracts Harmonia 2014/14/M/ST2/00428, Opus 2014/13/B/ST2/02543, 2014/15/B/ST2/03998, and 2015/19/B/ST2/02861, Sonata-bis 2012/07/E/ST2/01406; the National Priorities Research Program by Qatar National Research Fund; the Ministry of Science and Education, grant no. 3.2989.2017 (Russia); the Programa Estatal de Fomento de la Investigaci{\'o}n Cient{\'i}fica y T{\'e}cnica de Excelencia Mar\'{\i}a de Maeztu, grant MDM-2015-0509 and the Programa Severo Ochoa del Principado de Asturias; the Thalis and Aristeia programs cofinanced by EU-ESF and the Greek NSRF; the Rachadapisek Sompot Fund for Postdoctoral Fellowship, Chulalongkorn University and the Chulalongkorn Academic into Its 2nd Century Project Advancement Project (Thailand); the Nvidia Corporation; the Welch Foundation, contract C-1845; and the Weston Havens Foundation (USA). \end{acknowledgments}

\bibliography{auto_generated}  
\cleardoublepage \appendix\section{The CMS Collaboration \label{app:collab}}\begin{sloppypar}\hyphenpenalty=5000\widowpenalty=500\clubpenalty=5000\vskip\cmsinstskip
\textbf{Yerevan Physics Institute, Yerevan, Armenia}\\*[0pt]
A.M.~Sirunyan$^{\textrm{\dag}}$, A.~Tumasyan
\vskip\cmsinstskip
\textbf{Institut f\"{u}r Hochenergiephysik, Wien, Austria}\\*[0pt]
W.~Adam, F.~Ambrogi, T.~Bergauer, M.~Dragicevic, J.~Er\"{o}, A.~Escalante~Del~Valle, M.~Flechl, R.~Fr\"{u}hwirth\cmsAuthorMark{1}, M.~Jeitler\cmsAuthorMark{1}, N.~Krammer, I.~Kr\"{a}tschmer, D.~Liko, T.~Madlener, I.~Mikulec, N.~Rad, J.~Schieck\cmsAuthorMark{1}, R.~Sch\"{o}fbeck, M.~Spanring, W.~Waltenberger, C.-E.~Wulz\cmsAuthorMark{1}, M.~Zarucki
% ...
\vskip\cmsinstskip
\dag: Deceased\\
1:  Also at Vienna University of Technology, Vienna, Austria\\
% ...
\end{sloppypar}
\end{document}